\documentclass[journal=jpclfh]{achemso}
\usepackage{graphicx}
\makeatletter
\let\saved@includegraphics\includegraphics
\AtBeginDocument{\let\includegraphics\saved@includegraphics}
\renewenvironment*{figure}{\@float{figure}}{\end@float}
\makeatother
\usepackage[utf8]{inputenc}
\usepackage[T1]{fontenc}
\usepackage{multirow}
\usepackage{subfigure}
\usepackage{mathtools}
\usepackage[colorlinks=true,linkcolor=blue,citecolor=black,urlcolor=blue]{hyperref}
\usepackage[labelfont=bf]{caption}
\usepackage[figurename=Figure]{caption}
\usepackage{siunitx}
\usepackage[dvipsnames]{xcolor}
\usepackage{soul}
\usepackage{xfrac}

\usepackage{siunitx}
\usepackage{soul,xcolor}

\usepackage{braket}
\usepackage{bm}
\usepackage{upgreek}
\usepackage{lineno}

\DeclareSIUnit\rydberg{Ry}
\DeclareSIUnit\angstrom{\text {Å}}
\usepackage{amsmath} % or simply amstext
\newcommand{\angstrom}{\text{\normalfont\AA}}
\newcommand{\iu}{\mathrm{i}}

\usepackage{titlecaps}
\Addlcwords{a, an, the, of, in, and, for, to, with}
\graphicspath{{./figures/}}

\author{Sufyan Shehada}
\email{s.shehada@fz-juelich.de}
\affiliation{Peter Gr\"{u}nberg Institut, Forschungszentrum J\"{u}lich and JARA, D-52425 J\"{u}lich, Germany}
\alsoaffiliation{RWTH Aachen University, 52056 Aachen, Germany}
\alsoaffiliation{Department of Physics, Arab American University, 240 Jenin, Palestine}

\author{Manuel dos Santos Dias}
\affiliation{Peter Gr\"{u}nberg Institut, Forschungszentrum J\"{u}lich and JARA, D-52425 J\"{u}lich, Germany}
\alsoaffiliation{Faculty of Physics, University of Duisburg-Essen and CENIDE, 47053 Duisburg, Germany}
\alsoaffiliation{Scientific Computing Department, STFC Daresbury Laboratory, Warrington WA4 4AD, United Kingdom}

\author{Muayad Abusaa}
\affiliation{Department of Physics, Arab American University, 240 Jenin, Palestine}

\author{Samir Lounis}
\email{s.lounis@fz-juelich.de}
\affiliation{Peter Gr\"{u}nberg Institut, Forschungszentrum J\"{u}lich and JARA, D-52425 J\"{u}lich, Germany}
\alsoaffiliation{Faculty of Physics, University of Duisburg-Essen and CENIDE, 47053 Duisburg, Germany}

\title{3$d$-oxide molecules to tailor  large magnetic anisotropy energies on MgO films}

\begin{document}

\begin{abstract}

Designing systems with large magnetic anisotropy energy (MAE) is desirable and critical for nanoscale magnetic devices. 
 A recent breakthrough achieved the theoretical limit of the MAE for 3$d$ transition metal atoms by placing a single Co atom on a MgO(100) surface, a result not replicated by standard first-principles simulations. Our study, incorporating Hubbard-$U$ correction and spin-orbit coupling, successfully reproduces and explains the high MAE of a Co adatom on a MgO (001) surface. We go further by exploring ways to enhance MAE in 3d transition metal adatoms through different structural geometries of 3d--O molecules on MgO. One promising structure, with molecules perpendicular to the surface, enhances MAE while reducing substrate interaction, minimizing spin fluctuations, and boosting magnetic stability. Additionally, we demonstrate significant control over MAE by precisely placing 3d--O molecules on the substrate at the atomic level.

\end{abstract}

\section{Introduction}

Surface-embedded molecular magnetic structures are of tremendous interest, as they represent the smallest magnetic units at the ultimate atomic scale~\cite{Khajetoorians2011,Khajetoorians2013,Loth2012}.  
Recent studies on magnetic adatoms with sizeable magnetic anisotropy energy (MAE) has been intense due to their potential applications in high-density information storage and quantum spin processing~\cite{gambardella2003giant,Hirjibehedin2007,PhysRevB.81.104426,PhysRevLett.111.236801,hu2014giant,beljakov2014spin,PhysRevLett.103.187201,Rau2014,khajetoorians2014hitting,Baumann2015}.
For instance, single Co atoms deposited onto a Pt (111) surface give rise to an MAE of about 9 meV per atom favoring an out-of-plane orientation of the magnetic moment, and the assembled Co nanoparticles have an MAE that is reliant on the coordination of a single atom~\cite{gambardella2003giant} while being enhanced by the polarization of the substrate~\cite{Bouhassoune2016}. If the Co atoms are separated from the Pt surface by graphene, the MAE is maintained at a significant value (MAE =-8.1 meV, where the minus sign indicates an out-of-plane MAE)~\cite{PhysRevLett.111.236801}. 
Ab-initio simulations predicted the possibility of achieving remarkable MAEs for Co or Ir dimers on graphene (MAE = -60 meV)~\cite{hu2014giant}, for Os adatoms on graphene nanoflakes (MAE = -22 meV)~\cite{beljakov2014spin}, and for Co dimers on benzene (MAE = -100 meV)~\cite{PhysRevLett.103.187201}. 
A substantial out-of-plane MAE would generate an energy barrier that could protect the magnetization from thermal or quantum fluctuations~\cite{ibanez2016zero,ibanez2018spin,bouaziz2020zero}, making it robust and stable and allowing the magnetization to be orientated in a preferred spatial direction for a sufficient duration of time, which would be practical for the realization of a magnetic bit.

Strategies for enhancing the MAE of magnetic adatoms are based on three vital aspects: a large spin-orbit coupling (SOC) energy, a significant orbital moment, and a special ligand field~\cite{Rau2014,khajetoorians2014hitting,ibanez2016zero,Donati2016,donati2021correlation}.
As a ligand field frequently quenches or reduces an orbital moment, by enforcing orbital degeneracies, it is difficult to attain a massive MAE without a suitable surface or substrate.
Recently, thin insulating layers of MgO developed into an appealing substrate for exploring various magnetic aspects pertaining to magnetic adatoms and molecules~\cite{Rau2014,Baumann2015a,Baumann2015,Paul2017,Yang2017,Donati2016,natterer2017reading,PhysRevB.100.180405,choi2017atomic,yang2021probing,willke2019magnetic,Willke2018a,yang2018electrically,shehada2021trends,shehada2022interplay,kim2022anisotropic,farinacci2022experimental,zhang2022electron,willke2021coherent,kim2021spin,PhysRevB.103.155405,singha2021mapping,singha2021engineering,Yang2019a,kovarik2022electron,garai2023microscopic}.

\begin{figure}[ht!]
    \centering
    \includegraphics[width=\textwidth]{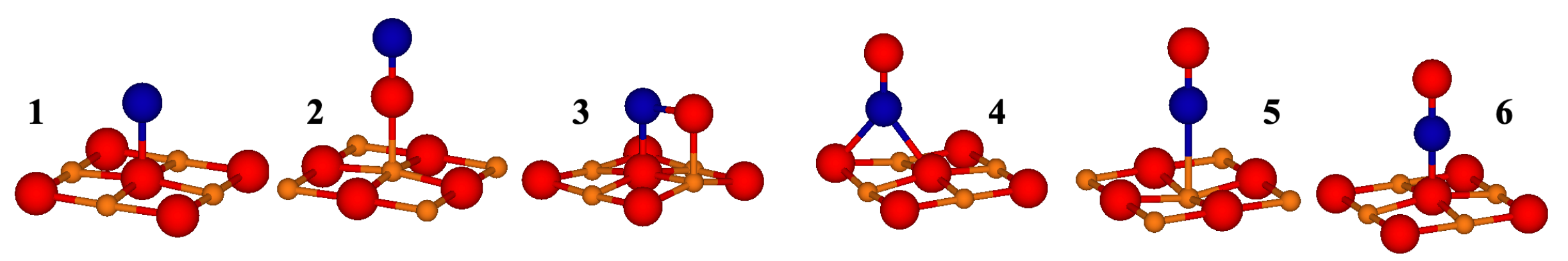}\vspace{-1em}
    \caption{Investigated atomic structures for $3d$ adatom and $3d$--O molecules on a bilayer of MgO. The different structures are numbered depending on the orientation of the molecule with respect to the substrate. 
    $3d$ atoms are represented by blue spheres, O by red spheres and Mg by orange spheres.}
    \label{fig:3dO_on_2MgO}
\end{figure}

On that very substrate, Rau et al.~\cite{Rau2014}  discovered  that a Co adatom adsorbing on the oxygen-top position of the MgO (001) surface (see Fig.~\ref{fig:3dO_on_2MgO}-1) is characterized by a large MAE since the underlying measured zero-field splitting with inelastic scanning tunneling spectroscopy (STS) reaches approximately 60 meV~\cite{Rau2014}. Assuming a spin $\frac{3}{2}$ for the  Co adatom implied  a MAE of the order of -90 meV~\cite{ou2015giant}, which broke records and reaches the magnetic anisotropy limit of $3d$ transition metals.
Details of the interaction between  Co and MgO surface, for instance the Co--O bond, determines  uniquely the underlying magnetic properties.
Theoretically, conventional density functional theory (DFT) simulations based on the local spin-density approximation (LSDA) or generalized gradient approximation (GGA) calculations do not recover the large MAE  of Co on MgO. Ou et al.~\cite{ou2015giant} predict that orbital-reordering  as described within DFT + $U$ can occur, which can enhance the MAE of Co on MgO while leading to gigantic MAEs for Ru (MAE = -110 meV) and Os on the same surface (MAE = -208 meV).

In this work, we investigate from ab-initio (see Method section) the remarkable MAE of Co, in particular, and the underlying  electronic mechanisms leading to its large value. Furthermore, we explore the MAE of the whole series of 3$d$ adatoms on MgO surface with the aim of identifying the ideal structural scenario for enhancing their MAE by forming molecules with an additional O atom (XO molecules, X being a 3$d$ atom), see Fig.~\ref{fig:3dO_on_2MgO}. Of our particular interest is the case where the molecule is perpendicular to the MgO surface, such that the interaction of Co with the substrate is minimized as shown in Fig.~\ref{fig:3dO_on_2MgO}-2,  which should reduce the substrate-induced spin fluctuations to a minimum~\cite{ibanez2016zero}.

\section{MAE of a single Co adatom on the bilayer of MgO}\label{MAE_single_Co}

 To set the stage for our study, we investigate the MAE of a single Co adatom on a bilayer of MgO surface. We consider the adatom on O-top site (see Fig.~\ref{fig:3dO_on_2MgO}-1) since it is the most stable adsorption site~\cite{Rau2014,shehada2021trends}. Regular LSDA calculations lead to a weak MAE (-10 meV), which favors an out-of-plane orientation of the magnetic moment. Since electronic correlations are expected to be crucial on MgO, we incorporated different values of  the Hubbard $U$ (and exchange parameter $J$) including SOC. Once correlations taken into account the MAE experiences a large increase and reaches the order of magnitude  experimental values for $U = 4, J = 1$ eV (MAE = -145.2 meV), and for $U = 6, J=1$ eV (-93.3 meV).

\begin{figure}[ht!]
    \centering
    \includegraphics[width=\textwidth]{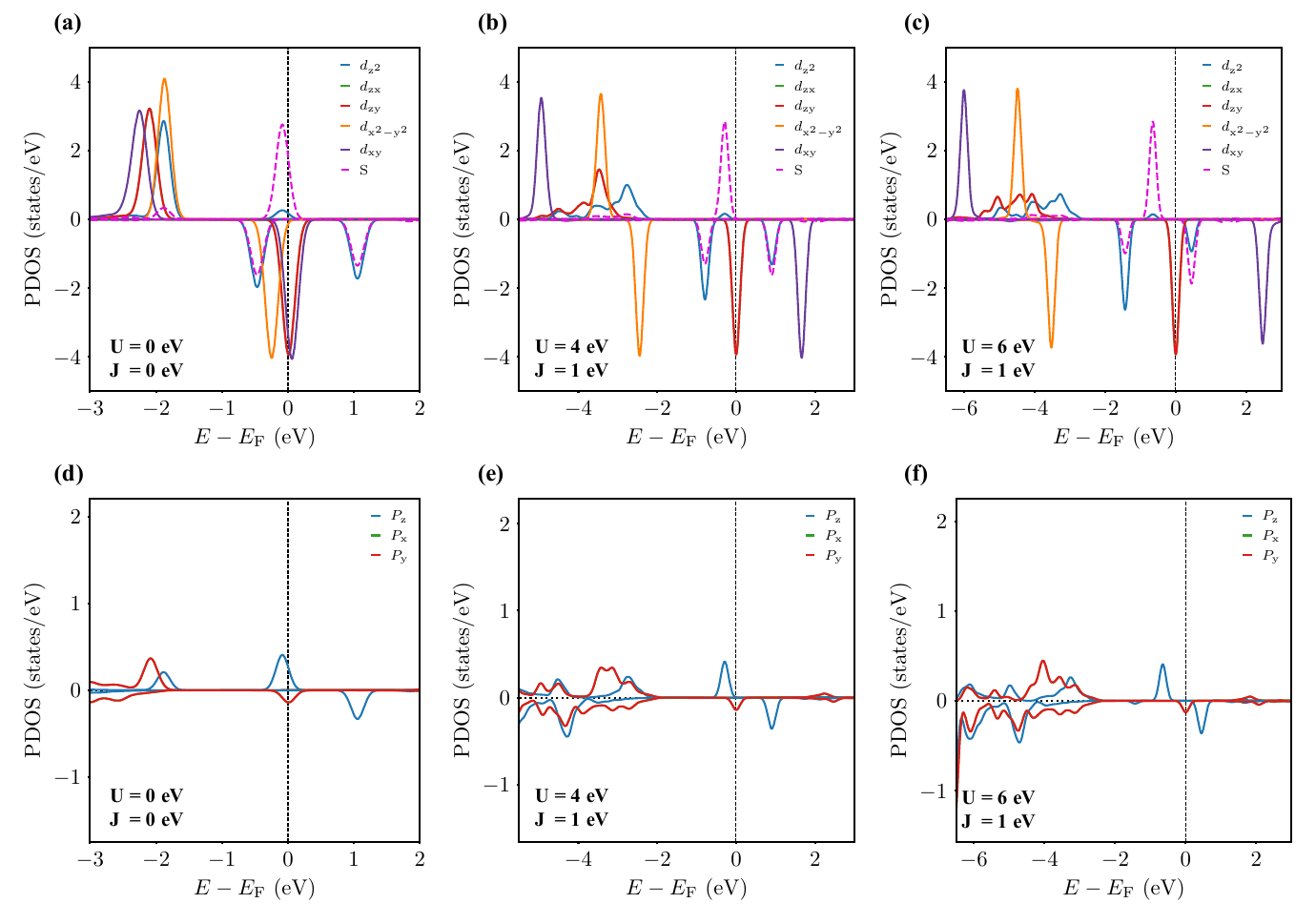}
    \caption{Effect of the Hubbard-$U$ correction on the electronic structure of a Co adatom (a, b, and c) atop oxygen (d, e, and f) from the MgO bilayer of MgO in the absence of SOC.
    (a, d) PDOS for $U = 0$ and $J=0$,
    (b, e) PDOS for $U = 4$ eV and $J=1$ eV,
    (c, f) PDOS for $U = 6$ eV and $J=1$ eV. The Fermi energy is marked by a vertical dashed line.}
    \label{fig:PDOS}
\end{figure}

To understand the undelying physics, we plot the corresponding partial density of states (PDOS) of Co adatom without including SOC (Fig.~\ref{fig:PDOS}).
A common feature in all investigated cases is the degeneracy at the Fermi level of the $d_{xz}$ and $d_{yz}$ states. The rest of the states experience a clear shifts with respect to the Fermi energy as soon as the Hubbard-$U$ correction included, which presumably trigger the aforementioned differences in the MAE. In the following we utilize degenerate first-order and non-degenerate second-order perturbation theories~\cite{wang1993first,brooks1940ferromagnetic,bruno1989tight,dai2008effects} in order to unveil the mechanisms shaping the large MAE characterizing the Co adatom.

% A common feature in all investigated cases is the degeneracy at the Fermi level of the $d_{xz}$ and $d_{yz}$ states. The rest of the states experience a clear shifts with respect to the Fermi energy as soon as the Hubbard-$U$ correction included, which presumably trigger the aforementioned differences in the MAE.\\

\textbf{First order degenerate perturbation theory.} In our discussion, first, we focus on the case of $U = 0$ eV and $J=0$ eV. We start by looking at the PDOS for the $d$-states in Fig.~\ref{fig:PDOS}-a. As aforementioned, 
the $d_{xz}$ and $d_{yz}$ minority-spin states are degenerate at the Fermi level for which we have to proceed with first-order degenerate perturbation theory~\cite{dai2008effects,sakurai1995modern,lu2013magnetic,shao2014carrier} in order to predict the impact of spin-orbit coupling. The $z$-component of the orbital momentum operator is the only one connecting both orbitals, implying that the SOC term of interest is  $\frac{1}{2}\xi \sigma_z\cdot \hat{\mathbf{L}}_z$, which simplifies to $-\frac{1}{2}\xi \hat{{L}}_z$ since the the two states are of minority-spin character. Here $\xi$ is the radial integral of the SOC with the associated atomic wave functions. This also means  that there is gain in energy  only when the moment points along the z-direction. 
We diagonalize the degenerate subuspace including the SOC term:

\begin{equation}\label{MSOC}
 \begin{bmatrix} \hat{\text{H}}_{11} & \hat{\text{H}}_{12} \\ \hat{\text{H}}_{21} & \hat{\text{H}}_{22} \end{bmatrix}
 %\,\,=\,\,
 % \begin{bmatrix}
 %  0 &
 %  \hat{\text{H}}_{12} \\
 %  \hat{\text{H}}_{21} &
 %  0 
 %  \end{bmatrix}
\,\,=\,\,-\frac{1}{2}\xi\,\,
   \begin{bmatrix}
    0 &
   \bra{d_{xz}} \hat{{L}_z}  \ket{d_{yz}} \\
   \bra{d_{yz}}  \hat{{L}_z}  \ket{d_{xz}} &
   0 
   \end{bmatrix}
\,\,=\,\,-\frac{1}{2}\xi\,\,
%\,\,=\,\,\xi(r)\cos\theta\,\,
   \begin{bmatrix}
    0 &     -\iu \\
    \iu &     0 
   \end{bmatrix}\,,
\end{equation}
and find as eingenvalues 
\begin{equation}\label{MAE1}
    \Delta \text{E}_{\pm} = \mp \frac{1}{2}\xi\,\,.
\end{equation}

The corresponding eigenstates are: ($d^\downarrow_1 = \frac{1}{\sqrt{2}}(d_{xz}+i\,d_{yz})$) for  eigenvalue $\text{E}_{1} = - \frac{1}{2}\xi$, which carries an orbital moment of 1 $\mu_B$ and ($d^\downarrow_2 = \frac{1}{\sqrt{2}}(d_{xz}-i\,d_{yz})$) for eigenvalue $\text{E}_{2} = \frac{1}{2}\xi$ with an orbital moment of -1 $\mu_B$.  
The electron initially shared by both orbitals $d_{xz}$ and $d_{yz}$ located at the Fermi energy will be located in the lowest energy state associated to $d^\downarrow_1$ just below the Fermi energy while the $d^\downarrow_2$ state becomes unoccupied. 

%%%%\samir{add a word regarding the orbital moment.
%$\bra{d_1^{\downarrow}} \hat{{L}_z}  \ket{d_1^{\downarrow}} = \frac{1}{2} i \left(\bra{d_{xz}} \hat{{L}_z}  \ket{d_{yz}} - \bra{d_{yz}} \hat{{L}_z}  \ket{d_{xz}} \right)=\frac{1}{2} i (-i-i)= 1$; $\bra{d_2^{\downarrow}} \hat{{L}_z}  \ket{d_2^{\downarrow}} = -\frac{1}{2} i \left(\bra{d_{xz}} \hat{{L}_z}  \ket{d_{yz}} - \bra{d_{yz}} \hat{{L}_z}  \ket{d_{xz}} \right)=-\frac{1}{2} i (-i-i)= -1$}

Overall, we conclude that the easy axis along z-direction is strongly favored by the degenerate states located at the Fermi energy, contributing to the MAE by a large value of 
%\new{ $-\frac{1}{2}\xi =-45$  meV, where we assumed that $\xi\approx 90$ meV for Co~\cite{daalderop1990first}}
$\text{MAE}_{1^\mathrm{st}\mathrm{order}}=-\frac{1}{2}\xi =-35$  meV, where we assumed that $\xi\approx 70$ meV for Co~\cite{van1991m2}. Obviously, it is the degeneracy of the $d_{xz}$ and $d_{yz}$ minority-spin states at the Fermi energy, which is responsible for the large out-of-plane MAE detected experimentally. The differences noticed among the simulations utilizing various values of $U$ and $J$ must be induced by the rest of the states, which are non-degenerate. These will be addressed in the following.\\

\textbf{Second-order non-degenerate perturbation theory.}

Here we evaluate the contributions to the MAE from the non-degenerate states. The 
MAE is determined by the matrix elements of SOC involving occupied and unoccupied states\cite{wang1993first}:
\begin{equation}\label{MAEeq}
    \text{MAE}_{2^\mathrm{nd}\mathrm{order}}= \frac{\xi^{2}}{4} \sum_{o,u,\sigma,\sigma^{'}} (1-2\delta_{\sigma \sigma^{'}}) \frac{\mid\bra{o^{\sigma}} \hat{{L}_z}\ket{u^{\sigma^{'}}}\mid^2-\mid\bra{o^{\sigma}} \hat{{L}_x}\ket{u^{\sigma^{'}}}\mid^2}{ \epsilon_{u,\sigma^{'}} -\epsilon_{o,\sigma}}
    \,\,,
\end{equation}
where $ o^{\sigma}$($u^{\sigma^{'}}$) and  $\epsilon_{o,\sigma}$($\epsilon_{u,\sigma^{'}}$) represent eigenstates and eigenvalues of occupied (unoccupied) states in spin state $\sigma$($\sigma^{'}$).
%The nonzero $\hat{{L}_z}$ and $\hat{{L}_x}$ matrix elements involving $d$-states are:
 % $\bra{d_{xz}} \hat{{L}_z}\ket{d_{yz}} = 1 \samir{-i}$,  $\bra{d_{x^2-y^2}} \hat{{L}_z}\ket{d_{xy}} = 2 \samir{-2i}$,  
 % $\bra{d_{z^2}} \hat{{L}_x}\ket{d_{xz},d_{yz}} = \sqrt{3}$\samir{$\mp i\sqrt{3}$}, 
 % $\bra{d_{xy}} \hat{{L}_x}\ket{d_{xz},d_{yz}} = 1$\samir{$\mp i$} and 
 % $\bra{d_{x^2-y^2}} \hat{{L}_x}\ket{d_{xz},d_{yz}} = 1$\samir{$i$}. \samir{Instead of the above and in order to avoid the problem that Manuel mentioned we could list the absolute values of the square of the matrix elements, which is not different from what would be obtained with the former results:}
 The nonzero $\hat{{L}_z}$ and $\hat{{L}_x}$ matrix elements involving $d$-states are:
  $\mid\bra{d_{xz}} \hat{{L}_z}\ket{d_{yz}}\mid^2 = 1$,  $\mid\bra{d_{x^2-y^2}} \hat{{L}_z}\ket{d_{xy}}\mid^2 = 4$,  
  $\mid\bra{d_{z^2}} \hat{{L}_x}\ket{d_{xz},d_{yz}}\mid^2 = 3$, 
  $\mid\bra{d_{xy}} \hat{{L}_x}\ket{d_{xz},d_{yz}}\mid^2 = 1$ and 
  $\mid\bra{d_{x^2-y^2}} \hat{{L}_x}\ket{d_{xz},d_{yz}} \mid^2= 1$.
Considering that all the majority-spin states are fully occupied and rather far away from the Fermi energy, as shown in Fig.~\ref{fig:PDOS}, the dominant contribution to the MAE can be attributed to the minority-spin states, spin-down occupied and spin-down unoccupied states,  $\sigma(\sigma^{'} = (\downarrow\downarrow)$. 
We neglect spin-flip contributions from spin-up occupied and spin-down unoccupied states for the qualitative analysis carried out in this section. In fact, there is a satellite majority-spin $d_{z^2}$ state showing up close to the Fermi energy. However, it emerges from the $p_z$ state of the underlying oxygen atom, which gives rise to a prominent s-state (shown as a dashed line in Fig.~\ref{fig:PDOS}).  
Eq.~\ref{MAEeq} is characterized by three primary finite dominant factors
\begin{equation}\label{MAEtot}
\text{MAE}_{2^\mathrm{nd}\mathrm{order}}=  -  \frac{\xi^{2}}{4}\,\,\frac{\mid\bra{d^\downarrow_{x^2-y^2}} \hat{{L}_z}\ket{d^\downarrow_{xy}}\mid^2}{\epsilon_{xy,^\downarrow} - \epsilon_{x^2-y^2,^\downarrow}}\,\, + \frac{\xi^{2}}{4}\,\,\frac{\mid\bra{d^\downarrow_{1}} \hat{{L}_x}\ket{d^\downarrow_{z^2}}\mid^2}{\epsilon_{z^2,^\downarrow} - \epsilon_{1,^\downarrow}}\,\,+ \frac{\xi^{2}}{4}\,\,\frac{\mid\bra{d^\downarrow_{1}} \hat{{L}_x}\ket{d^\downarrow_{xy}}\mid^2}{\epsilon_{xy,^\downarrow} - \epsilon_{1,^\downarrow}}\,\,,
\end{equation}
which are listed in Table~\ref{table:MAEALL} for different values of $U$ and $J$ and compared to the value obtained from first-order degenerate perturbation theory. 
Since static correlations increase the energy splitting between the occupied $d^\downarrow_{x^2-y^2}$ and unoccupied $d^\downarrow_{xy}$, the first term in Eq.~\ref{MAEeq}, favoring the out-of-plane easy-axis, reduces in magnitude. The same trend is followed by the third term, which however favors an in-plane orientation of the moment in contrast to the second term. Clearly, there is a complex competition between the different terms, which imposes a  reduction of the MAE emerging from first-order degenerate perturbation theory.

\begin{table}[ht!]
\centering
\begin{tabular}{ c |c |c| c }
 \hline
 \hline
  MAE & $U$ =0, $J$ =0 & $U$ =4, $J$ =1 & U=6, $J$ =1 \\ [.3cm]
 \hline
  $- \frac{\xi^{2}}{4}\,\,\frac{\mid\bra{d^\downarrow_{x^2-y^2}} \hat{{L}_z}\ket{d^\downarrow_{xy}}\mid^2}{\epsilon_{xy,^\downarrow} - \epsilon_{x^2-y^2,^\downarrow}}$ & -15.75 & -1.19 & -0.82  \\ [.3cm]
 \hline
 $+ \frac{\xi^{2}}{4}\,\,\frac{\mid\bra{d^\downarrow_{1}} \hat{{L}_x}\ket{d^\downarrow_{z^2}}\mid^2}{\epsilon_{z^2,^\downarrow} - \epsilon_{1,^\downarrow}}$ & 6.97 & 8.13 & 16.73  \\ [.3cm]
 \hline
 $+ \frac{\xi^{2}}{4}\,\,\frac{\mid\bra{d^\downarrow_{1}} \hat{{L}_x}\ket{d^\downarrow_{xy}}\mid^2}{\epsilon_{xy,^\downarrow} - \epsilon_{1,^\downarrow}}$ & 38.89 & 1.47 & 1.00 \\ [.3cm]
 \hline 
 $\text{MAE}_{1^\mathrm{st}\mathrm{order}} = - \frac{1}{2}\xi \,\,$ & -35.0 &  -35.0 & -35.0\\ [.3cm]
 \hline
$\text{MAE}_{1^\mathrm{st}\mathrm{order}} + \text{MAE}_{2^\mathrm{nd}\mathrm{order}} $ & -4.90 & -26.59 & -18.08
 \\ [.3cm]
 \hline
 \hline
\end{tabular}
\caption{Various contributions from first-order degenerate and second-order non-degenerate perturbation theories to the MAE of Co adatom on the bilayer of MgO in the oxygen-top position from DFT + $U$  with the absence of SOC. A negative sign of the MAE favors an out-of-plane magnetization. $U$ and $J$ are given in eV, while the MAE is in meV.}
\label{table:MAEALL}
\end{table}
After summing up the different contributions to the MAE as obtained from first- and second-order perturbation theory, we recover qualitatively the trends found from the full ab-initio calculations as reported in Table~\ref{table:MAEALL}. %\sufyan{or Table~\ref{table:MAEALL_xi=70}}.
This shows that the main mechanism favoring the out-of-plane orientation of the magnetic moment with a large MAE is driven by the SOC-induced lifting of the degeneracy of the minority-spin $d_{xz}$ and $d_{yz}$ states located at the Fermi energy. 
We note that the magnitude of the predicted MAE is of the same order than the one obtained by  a previous theoretical study~\cite{ou2015giant} based on full-potential augmented plane wave calculations. Their argument, however, to explain the large MAE of Co adatom is different from ours and is based on a non-trivial reordering of the occupation matrix generated with static correlations.

In the next section, we  investigate the MAE of the rest of the 3$d$ series of adatoms deposited on MgO and explore the possibility of enhancing their MAE by considering 3$d$--O molecules as potential adsorbates  (see Fig.~\ref{fig:3dO_on_2MgO}). In the following we limit our simulations incorporating correlations to the case of $U = 6$ and $J = 1$ eV since the MAE value obtained for a Co adatom is the closest to the experimentally measured one~\cite{Rau2014}.

\section{MAE of 3$d$ adatoms and 3$d$--O molecules on the bilayer of MgO}

\begin{figure}[ht!]
    \centering
    \includegraphics[width=\textwidth]{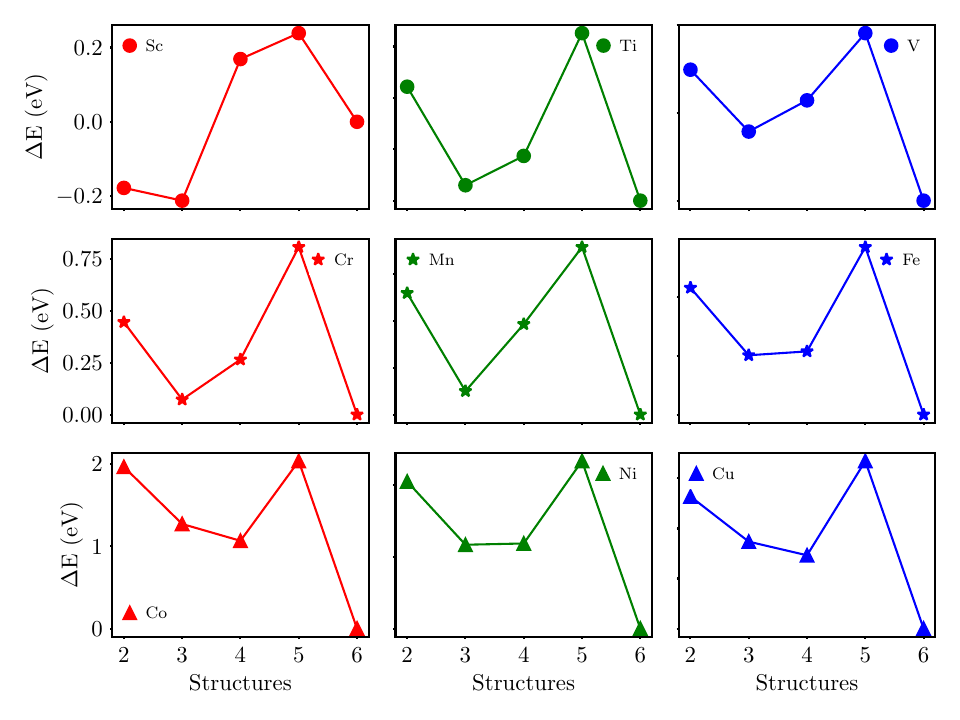}\vspace{-1em}
    \caption{Total energy difference of  $3d$--O molecules on the bilayer of MgO with respect to the one of structure 6. The structure number indicates a particular nanostructure illustrated in Fig.~\ref{fig:3dO_on_2MgO}. }
    \label{fig:3dO_on_2}
\end{figure}

Here, we address the main topic of our investigation, namely $3d$ adatoms on the bilayer of MgO and the $3d$--O molecules placed on the MgO bilayer considering different structures.
After structural relaxation, using LSDA + $U$ ($U$ = 6 eV and $J$ = 1 eV) total energy calculations, we classified the results into six structures shown in Fig.~\ref{fig:3dO_on_2MgO}.
As illustrated in Fig.~\ref{fig:3dO_on_2}, structure 6 is energetically the most favorable one for all $3d$--O molecules except for the Sc--O molecule case, while structure 5 is the one that is the least favorable. 
Independent from their relative stability, we study the MAE of all the converged nanostructures. 
\\
Most of the investigated structures have C$_{4V}$ symmetry except for structures 3 and 4. For the latter cases, we explore two in-plane rotations of the magnetization and calculate the MAE by taking the energy difference $\text{MAE}= \text{E}^{\hat{z}} -\text{E}^{\phi}$, 
where $\phi$ is the azimuthal angle. The cases where the magnetic moment points in-plane with an azimuthal angle $\phi = 0^\circ$, $\phi$ = 45$^{\circ}$ and $\phi$ = 90$^{\circ}$ are  respectively denoted as  a, b and c.

\begin{figure}[ht!]
    \centering
     %\textbf{Your title}
    \includegraphics[width=\textwidth]{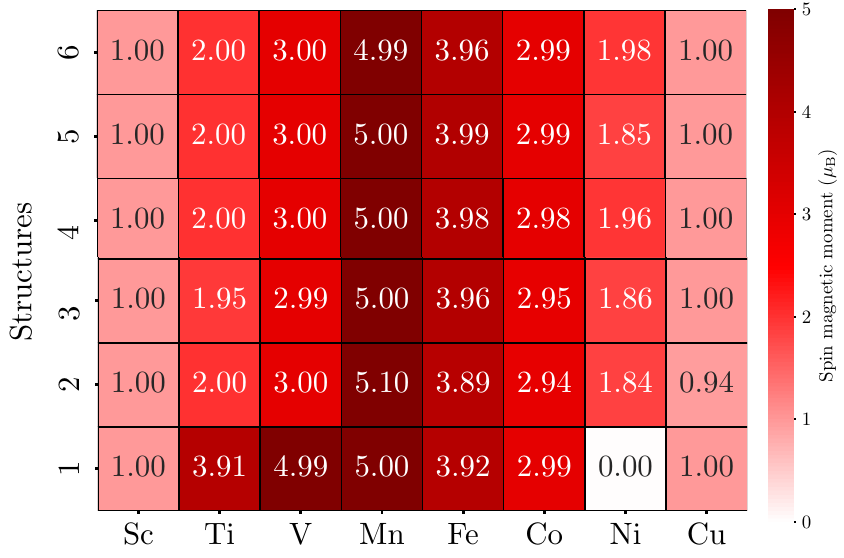}\vspace{-1em}
    \caption{Spin magnetic moments of $3$d adatoms and  $3d$--O molecules on a bilayer of MgO. The structure number indicates a particular nanostructure with a specific orientation of the magnetization as illustrated in Fig.~\ref{fig:3dO_on_2MgO}. }
    \label{fig:3dO_spin_2MgO}
\end{figure}

It is valuable to explore how the magnetic moments of the 3$d$ atoms are changed once embedded in the 3$d$--O molecules as summarized in Fig.~\ref{fig:3dO_spin_2MgO}. Among all investigated nanostructures, only three single adatoms do not follow Hund's first rule: Ni, Ti and V. 
The spin moments of the different deposited nanostructures  are generally unaffected by the positions of the  $3d$--O molecules and the electronic occupation is consistent with a nominal valence of [Ar]$4s^23d^n$. We note that Cr is not considered in the current work due to computational difficulties to reach self-consistency. Its MAE is expected to be nevertheless negligible.

%\begin{figure}[ht!]
%    \centering
%    \includegraphics[width=\textwidth]{Figures/6.4_new.pdf}\vspace{-1em}
%   \caption{Magnetization directions along [100], [110], and [010] corresponding to $\phi$ = 0$^{\circ}$, 45$^{\circ}$, and 90$^{\circ}$, respectively, for $3d$ adatoms and $3d$--O molecules on a bilayer of MgO. 
 %   $3d$ atoms are represented by blue spheres, O by red spheres and Mg by orange spheres.}
  %  \label{fig:3dO_MAE_2MgO}
%\end{figure}

\begin{figure}[ht!]
    \centering
     %\textbf{Your title}
    \includegraphics[width=\textwidth]{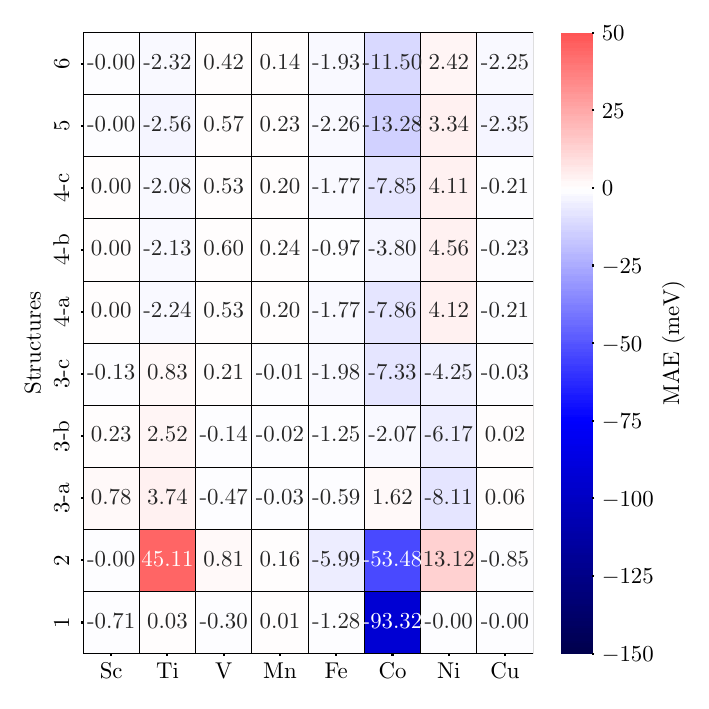}\vspace{-1em}
    \caption{MAE of $3$d adatoms and  $3d$--O molecules on a bilayer of MgO. The structure number indicates a particular nanostructure with a specific orientation of the magnetization as illustrated in Fig.~\ref{fig:3dO_on_2MgO}. Configurations a, b, and c correspond to the energy differences evaluated when the moment is rotated in-plane with with the azimuthal angles $\phi = 0^\circ$, $\phi = 45^\circ$ and $\phi = 90^\circ$, respectively. A negative sign of the MAE favors an out-of-plane magnetization. }
    \label{fig:MAE_all_results}
\end{figure}

The calculated MAEs are presented  in Fig.~\ref{fig:MAE_all_results}.
Ti, Fe, Co, and Ni structures generally yield significant MAE ranging from a few to tens meV. At the same time, the MAE values of Sc, V, Mn, and Cu are close to zero for all structures except the Cu--O molecule, which is characterized by an out-of-plane MAE of 2.35 and 2.25 meV in structures 5 and 6, respectively.

Among all the $3$d adatoms and $3d$--O molecules, a single Co adatom on the bilayer of MgO in the oxygen-top position exhibits the most significant MAE values (an out-of-plane MAE of -93 meV), representing the magnetic anisotropy limit of $3$d adatoms and $3d$--O molecules on MgO as obtained from our simulations.
Moreover, Co--O molecule in structure 2 (see Fig.~\ref{fig:3dO_on_2MgO}-2) is perpendicular to the substrate and has the second largest out-of-plane MAE of 53 meV compared to the rest of   explored nanostructures, see Fig.~\ref{fig:MAE_all_results}. 
A closer look at the Ti MAE values in Fig.~\ref{fig:MAE_all_results} reveals that the perpendicular Ti--O molecule in structure 2 has the largest in-plane MAE of 45 meV, with the remaining structures MAE values significantly decreasing (range between 0 and 4 meV). 
Except for structure 3-a, Co--O molecules prefer all an out-of-plane orientation of the magnetization.

We expect structure 2, to be the one where the 3$d$ atoms are less interacting with the substrate, which should favor magnetic stability if allowed by the right out-of-plane MAE. In principle its MAE should be close to the free standing molecule. 
To examine this scenario, we calculated the MAE of the isolated Co--O molecule for different bond lengths between Co and O atoms (d$_\text{Co--O}$), see Table~\ref{table:MAE_free_U_6}, to cover all the bond lengths of Co--O molecules on the bilayer of MgO (structures 2 to 6 in Fig.~\ref{fig:3dO_on_2MgO}).
Table~\ref{table:MAE_free_U_6} shows an excellent agreement between the computed MAE of the isolated Co--O molecule and the perpendicular Co--O molecule on the MgO bilayer, which confirms  our expectations.  
Moreover, Table~\ref{table:MAE_free_U_6} reveals a minimal effect of the bond length on the computed MAEs of the isolated Co--O molecule.

\begin{table}[ht!]
\centering
\begin{tabular}{|c| c| c| c |c|c|}
 \hline
 d$_\text{Co--O}$ ( \angstrom )  & 1.7 & 1.8 & 1.9 & 2  &CoO/2MgO\\
 \hline
 MAE (meV) & -46.06 & -48.25 & -50.89 & -52.11 & -53.48 \\
 \hline
\end{tabular}
%\end{ruledtabular}
\caption{\label{table:MAE_free_U_6} MAE of the isolated Co--O molecule for different bond lengths between Co and O atoms (d$_\text{Co--O}$).
CoO/2MgO is the perpendicular Co--O molecule on the MgO bilayer (see Fig.~\ref{fig:3dO_on_2MgO}-2). A negative sign of the MAE favors an out-of-plane magnetization. Here we used $U = 6$ and $J =1$ eV.}
\end{table}

\begin{figure}[ht!]
    \centering
     %\textbf{Your title}
    \includegraphics[width=\textwidth]{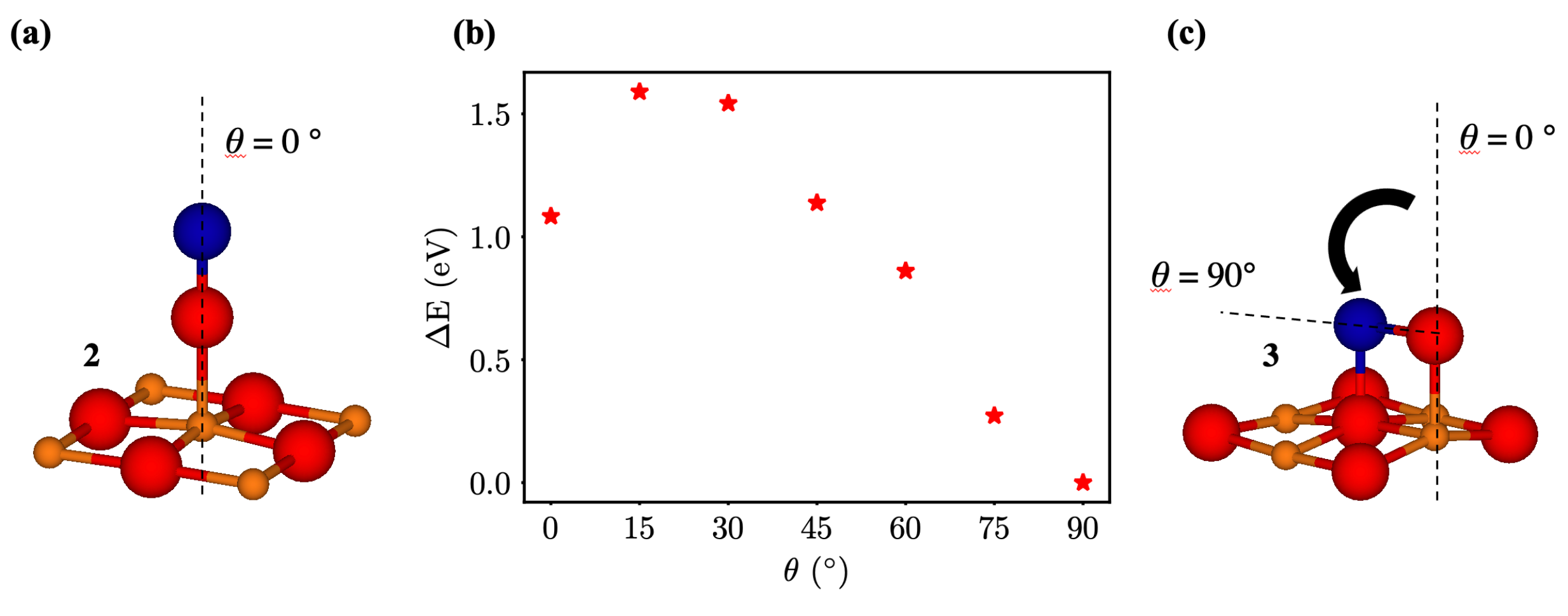}\vspace{-1em}
    \caption{Energy barrier for the metastable perpendicular Co--O molecule on the bilayer of MgO.
     (a) the perpendicular Co--O molecule on the MgO bilayer ($\theta= 0^{\circ}$), (b) total energy difference of Co--O molecule on the bilayer of MgO with respect to structure 3, as a function of the rotation  angle ($\theta$) and (c) the horizontal Co--O molecule on the MgO bilayer ($\theta= 90^{\circ}$).
     $\theta$ is the  rotation angle away from the $z$-axis towards the $x$-axis. Here we used $U = 6$ and $J =1$ eV.}
    \label{fig:CoO2MgO}
\end{figure}

The perpendicular Co--O molecule  (structure 2 in Fig.~\ref{fig:3dO_on_2MgO}) with a large MAE is energetically less stable than the horizontal one (structure 3 in Fig.~\ref{fig:3dO_on_2MgO})  with a weaker MAE. However, our simulations indicate that the perpendicular molecule is metastable and and could be protected by an energy barrier, preventing it to fall into the horizontal configuration and would enable its experimental realization. Fig.~\ref{fig:CoO2MgO} shows the energy difference associated to the rotation of the molecule on MgO, where a large barrier  of 0.6 eV can be clearly recognized.

%\begin{figure}[ht!]
%    \centering
%     %\textbf{Your title}
%    \includegraphics[width=\textwidth]{Figures/6.7_new.pdf}\vspace{-1em}
%    \caption{Energy barrier for the metastable perpendicular Fe--O molecule on the bilayer of MgO.
%     (a) the perpendicular Fe--O molecule on the MgO bilayer ($\theta= 0^{\circ}$), (b) total energy difference of Fe--O molecule on the bilayer of MgO with respect to structure 3, as a function of the rotating angle ($\theta$) and (c) The horizontal Fe--O molecule on the MgO bilayer ($\theta= 90^{\circ}$).
%     $\theta$ is the  rotation angle away from the z-axis towards the x-axis. Here we used $U = 6$ and $J =1$ eV.}
%    \label{fig:FeO2MgO}
%\end{figure} 

\section*{Discussion and conclusion}

We presented the results of ab initio calculations on the MAE of  $3d$--O molecules free-standing or deposited on the MgO bilayer, which were compared to the case of $3d$ adatoms on the oxygen-top position of the MgO bilayer. We explored, in particular, their structural, electronic, and magnetic properties and scrutinized the impact of the existence of an extra oxygen atom attached to $3d$ adatoms on the MAE. The physics of the latter is mainly explained by applying degenerate and non-degenerate perturbation theories.

We evidenced the ability to substantially modify the MAE via atomic control of the location of the $3d$--O molecules on the bilayer of MgO substrate.
In particular, we revealed the possibility of having the  $3$d--O molecules perpendicular to the substrate with the 3$d$ adatom being atop the oxygen atom of the molecule, which should minimize spin-fluctuations triggered by the interaction with the substrate. These molecules can be characterized by large MAE similar to that of the isolated Co adatom. Both aspects, large MAE and weak coupling to the substrate are the right ingredients to enable magnetic stability of the nanostructure, which so far has not been achieved for 3$d$ adatoms. 
In fact, the perpendicular $3$d--O molecules on the bilayer of MgO act like the isolated  $3$d--O molecule indicating the weak impact of the substrate on the MAE,  especially in the cases of Co--O, Ti--O, Ni--O, and Fe--O perpendicular molecules.
Although the aforementioned perpendicular molecule is a metastable structure, it could be protected by an energy barrier, which makes its experimental realization via atomic manipulation with scanning tunneling microscopy  possible. 
Moreover, we evidenced the ability to substantially modify the MAE by atomic control by controlling the location of the $3d$--O molecules on the substrate.

\section*{Methods}

The simulations are conducted within density functional theory as implemented in the Quantum ESPRESSO code with scalar relativistic \cite{vanderbilt1990soft} as well as fully relativistic ultra-soft pseudopotentials (USPPs)\cite{dal2005spin,dal2014pseudopotentials}. We assume the local spin density approximation (LSDA)~\cite{PhysRevB.23.5048} and consider electronic correlations within the formulation of the Hubbard-$U$ correction based on Ref.~\cite{liechtenstein1995density}.

In our work, we discuss two types of simulations: $3d$ adatoms on the oxygen-top position of the MgO bilayer and $3d$--O molecules free-standing or deposited on the MgO bilayer.
For the case of free-standing $3d$--O molecules, we employed cubic periodic cells with a lattice constant of \SI{20}{\angstrom}, in order to minimize interactions between periodic replicas of the dimers, and assumed $\Gamma$-point sampling of the Brillouin zone while using different bond lengths between $3d$ adatoms and O atoms (d$_\text{3d--O}$).

To accommodate the  3$d$ adatoms and 3$d$--O molecules on the bilayer of MgO,
we used the theoretical lattice constants of a bilayer of MgO obtained  from LDA (\SI{4.065}{\angstrom}).
We then set up $3\times3$ supercells such that the 3$d$ adatoms are deposited on top of the oxygen positions, as shown in (Fig.~\ref{fig:3dO_on_2MgO}-1).
For 3$d$--O molecules on bilayer of MgO, we set the 3$d$--O molecules on different structures as shown in (Fig.~\ref{fig:3dO_on_2MgO}).
The supercells contain 73 and 74 atoms in total for the case of $3d$ adatoms on the oxygen-top position and $3d$--O molecules deposited on the MgO bilayer MgO, respectively, and a vacuum thickness equivalent to 9 layers of MgO.
We adopted a $4\times4\times1$ k-mesh in both cases and the cell dimensions were kept fixed while all atomic positions were allowed to relax in $z$-axis.

The rotation of the magnetic moment of the free-standing Co--O molecules was studied by the constrained DFT approach explained in Ref.~\cite{PhysRevB.91.054420}. 
In order to ensure that the different magnetic states were comparable, for each fixed magnetic configuration 
we performed a sequence of self-consistent constrained calculations. We highlight that the MAE obtained for the free-standing molecules are quantitatively similar when utilizing USPP or PAW in either LSDA or the generalized gradient approximation (GGA),see Table~\ref{table:MAE_free}.

\begin{table}[ht!]
\centering
\begin{tabular}{|c| c| c| c |c|}
 \hline
  PPs & GGA PAW & GGA USPP & PAW & LSDA USPP \\
 \hline
MAE (meV) & -13.5   & -12.9 & -10.0  & -12.84   \\
 \hline
\end{tabular}
%\e
\caption{\label{table:MAE_free} MAE of the isolated Co--O molecule from DFT+SOC total energy calculations for different types of pseudopotentials (PPs) and exchange-correlation functionals, (d$_\text{Co-O} = 2 \angstrom$). A negative sign of the MAE favors an out-of-plane magnetization.}
\end{table}

\iffalse
\begin{table}[ht!]
\centering
\begin{tabular}{|c| c| c| c |c|}
 \hline
  PPs & GGA PAW & GGA USPP & PAW & LSDA USPP \\
 \hline
MAE (meV) & -13.5   & -12.9 & -10.0  & -12.84   \\
 \hline
\end{tabular}
%\e
\caption{\label{table:MAE_free} MAE of the isolated Co--O molecule from DFT+SOC total energy calculations for different types of pseudopotentials (PPs) and exchange-correlation functionals, (d$_\text{Co-O} = 2 \angstrom$). A negative sign of the MAE favors an out-of-plane magnetization.}
\end{table}

As an initial test, we computed the MAE for free-standing Co--O molecules with different types of fully relativistic pseudopotentials shown in Table.~\ref{table:MAE_free}. Since the results were quantitatively similar, we opted for LSDA+USPP for the rest of the MAE calculations due to the low computational cost, especially when we add Hubbard-$U$ correction. The MAE is defined as the difference in the total energies between the two magnetic states: in-plane magnetization along the x-axis and out-of-plane magnetization along the z-axis ($ \text{MAE}= \text{E}^{\hat{z}} -\text{E}^{\hat{x}}$). In our convention, a negative MAE corresponds to a favorable out-of-plane magnetization.
\fi

\subsection{Data availability} The data needed to evaluate the conclusions in the paper are present in the manuscript.
\subsection{Code availability} All codes used for this work are open-source. Quantum ESPRESSO can be found at \url{https:/www.quantum-espresso.org/download}.

\section{Acknowledgments}

This work was supported by the Federal Ministry of Education and Research of Germany in the framework
of the Palestinian-German Science Bridge (BMBF grant number 01DH16027).  
We acknowledge the computing time granted by the JARA-HPC Vergabegremium and VSR commission on the supercomputer JURECA at Forschungszentrum Jülich~\cite{jureca} and RWTH Aachen University under project p0020362.

\section{Author contributions}
S.L. initiated, designed and supervised the project. S.S. performed the simulations and post processed the data. S.SA., M.d.S.D, M.A., and S.L. discussed the results. S.S. and S.L. wrote the manuscript to which all co-authors contributed.

\section{Competing Interests}
The authors declare no competing interests.

\section{Correspondence} Correspondence and requests for materials should be addressed to S.S. (email: s.shehada@fz-juelich.de) or to S.L. (email: s.lounis@fz-juelich.de).

\bibliography{references}

\providecommand{\latin}[1]{#1}
\makeatletter
\providecommand{\doi}
  {\begingroup\let\do\@makeother\dospecials
  \catcode`\{=1 \catcode`\}=2 \doi@aux}
\providecommand{\doi@aux}[1]{\endgroup\texttt{#1}}
\makeatother
\providecommand*\mcitethebibliography{\thebibliography}
\csname @ifundefined\endcsname{endmcitethebibliography}  {\let\endmcitethebibliography\endthebibliography}{}
\begin{mcitethebibliography}{59}
\providecommand*\natexlab[1]{#1}
\providecommand*\mciteSetBstSublistMode[1]{}
\providecommand*\mciteSetBstMaxWidthForm[2]{}
\providecommand*\mciteBstWouldAddEndPuncttrue
  {\def\EndOfBibitem{\unskip.}}
\providecommand*\mciteBstWouldAddEndPunctfalse
  {\let\EndOfBibitem\relax}
\providecommand*\mciteSetBstMidEndSepPunct[3]{}
\providecommand*\mciteSetBstSublistLabelBeginEnd[3]{}
\providecommand*\EndOfBibitem{}
\mciteSetBstSublistMode{f}
\mciteSetBstMaxWidthForm{subitem}{(\alph{mcitesubitemcount})}
\mciteSetBstSublistLabelBeginEnd
  {\mcitemaxwidthsubitemform\space}
  {\relax}
  {\relax}

\bibitem[Khajetoorians \latin{et~al.}(2011)Khajetoorians, Wiebe, Chilian, and Wiesendanger]{Khajetoorians2011}
Khajetoorians,~A.~A.; Wiebe,~J.; Chilian,~B.; Wiesendanger,~R. Realizing all-spin--based logic operations atom by atom. \emph{Science} \textbf{2011}, \emph{332}, 1062--1064\relax
\mciteBstWouldAddEndPuncttrue
\mciteSetBstMidEndSepPunct{\mcitedefaultmidpunct}
{\mcitedefaultendpunct}{\mcitedefaultseppunct}\relax
\EndOfBibitem
\bibitem[Khajetoorians \latin{et~al.}(2013)Khajetoorians, Baxevanis, H{\"u}bner, Schlenk, Krause, Wehling, Lounis, Lichtenstein, Pfannkuche, Wiebe, and Wiesendanger]{Khajetoorians2013}
Khajetoorians,~A.~A.; Baxevanis,~B.; H{\"u}bner,~C.; Schlenk,~T.; Krause,~S.; Wehling,~T.~O.; Lounis,~S.; Lichtenstein,~A.; Pfannkuche,~D.; Wiebe,~J.; Wiesendanger,~R. Current-Driven Spin Dynamics of Artificially Constructed Quantum Magnets. \emph{Science} \textbf{2013}, \emph{339}, 55--59\relax
\mciteBstWouldAddEndPuncttrue
\mciteSetBstMidEndSepPunct{\mcitedefaultmidpunct}
{\mcitedefaultendpunct}{\mcitedefaultseppunct}\relax
\EndOfBibitem
\bibitem[Loth \latin{et~al.}(2012)Loth, Baumann, Lutz, Eigler, and Heinrich]{Loth2012}
Loth,~S.; Baumann,~S.; Lutz,~C.~P.; Eigler,~D.~M.; Heinrich,~A.~J. Bistability in Atomic-Scale Antiferromagnets. \emph{Science} \textbf{2012}, \emph{335}, 196--199\relax
\mciteBstWouldAddEndPuncttrue
\mciteSetBstMidEndSepPunct{\mcitedefaultmidpunct}
{\mcitedefaultendpunct}{\mcitedefaultseppunct}\relax
\EndOfBibitem
\bibitem[Gambardella \latin{et~al.}(2003)Gambardella, Rusponi, Veronese, Dhesi, Grazioli, Dallmeyer, Cabria, Zeller, Dederichs, Kern, \latin{et~al.} others]{gambardella2003giant}
Gambardella,~P.; Rusponi,~S.; Veronese,~M.; Dhesi,~S.; Grazioli,~C.; Dallmeyer,~A.; Cabria,~I.; Zeller,~R.; Dederichs,~P.; Kern,~K.; others Giant magnetic anisotropy of single cobalt atoms and nanoparticles. \emph{science} \textbf{2003}, \emph{300}, 1130--1133\relax
\mciteBstWouldAddEndPuncttrue
\mciteSetBstMidEndSepPunct{\mcitedefaultmidpunct}
{\mcitedefaultendpunct}{\mcitedefaultseppunct}\relax
\EndOfBibitem
\bibitem[Hirjibehedin \latin{et~al.}(2007)Hirjibehedin, Lin, Otte, Ternes, Lutz, Jones, and Heinrich]{Hirjibehedin2007}
Hirjibehedin,~C.~F.; Lin,~C.-Y.; Otte,~A.~F.; Ternes,~M.; Lutz,~C.~P.; Jones,~B.~A.; Heinrich,~A.~J. Large Magnetic Anisotropy of a Single Atomic Spin Embedded in a Surface Molecular Network. \emph{Science} \textbf{2007}, \emph{317}, 1199--1203\relax
\mciteBstWouldAddEndPuncttrue
\mciteSetBstMidEndSepPunct{\mcitedefaultmidpunct}
{\mcitedefaultendpunct}{\mcitedefaultseppunct}\relax
\EndOfBibitem
\bibitem[B\l{}o\ifmmode~\acute{n}\else \'{n}\fi{}ski \latin{et~al.}(2010)B\l{}o\ifmmode~\acute{n}\else \'{n}\fi{}ski, Lehnert, Dennler, Rusponi, Etzkorn, Moulas, Bencok, Gambardella, Brune, and Hafner]{PhysRevB.81.104426}
B\l{}o\ifmmode~\acute{n}\else \'{n}\fi{}ski,~P.; Lehnert,~A.; Dennler,~S.; Rusponi,~S.; Etzkorn,~M.; Moulas,~G.; Bencok,~P.; Gambardella,~P.; Brune,~H.; Hafner,~J. Magnetocrystalline anisotropy energy of Co and Fe adatoms on the (111) surfaces of Pd and Rh. \emph{Phys. Rev. B} \textbf{2010}, \emph{81}, 104426\relax
\mciteBstWouldAddEndPuncttrue
\mciteSetBstMidEndSepPunct{\mcitedefaultmidpunct}
{\mcitedefaultendpunct}{\mcitedefaultseppunct}\relax
\EndOfBibitem
\bibitem[Donati \latin{et~al.}(2013)Donati, Dubout, Aut\`es, Patthey, Calleja, Gambardella, Yazyev, and Brune]{PhysRevLett.111.236801}
Donati,~F.; Dubout,~Q.; Aut\`es,~G.; Patthey,~F.; Calleja,~F.; Gambardella,~P.; Yazyev,~O.~V.; Brune,~H. Magnetic Moment and Anisotropy of Individual Co Atoms on Graphene. \emph{Phys. Rev. Lett.} \textbf{2013}, \emph{111}, 236801\relax
\mciteBstWouldAddEndPuncttrue
\mciteSetBstMidEndSepPunct{\mcitedefaultmidpunct}
{\mcitedefaultendpunct}{\mcitedefaultseppunct}\relax
\EndOfBibitem
\bibitem[Hu and Wu(2014)Hu, and Wu]{hu2014giant}
Hu,~J.; Wu,~R. Giant magnetic anisotropy of transition-metal dimers on defected graphene. \emph{Nano letters} \textbf{2014}, \emph{14}, 1853--1858\relax
\mciteBstWouldAddEndPuncttrue
\mciteSetBstMidEndSepPunct{\mcitedefaultmidpunct}
{\mcitedefaultendpunct}{\mcitedefaultseppunct}\relax
\EndOfBibitem
\bibitem[Beljakov \latin{et~al.}(2014)Beljakov, Meded, Symalla, Fink, Shallcross, Ruben, and Wenzel]{beljakov2014spin}
Beljakov,~I.; Meded,~V.; Symalla,~F.; Fink,~K.; Shallcross,~S.; Ruben,~M.; Wenzel,~W. Spin-crossover and massive anisotropy switching of 5d transition metal atoms on graphene nanoflakes. \emph{Nano letters} \textbf{2014}, \emph{14}, 3364--3368\relax
\mciteBstWouldAddEndPuncttrue
\mciteSetBstMidEndSepPunct{\mcitedefaultmidpunct}
{\mcitedefaultendpunct}{\mcitedefaultseppunct}\relax
\EndOfBibitem
\bibitem[Xiao \latin{et~al.}(2009)Xiao, Fritsch, Kuz'min, Koepernik, Eschrig, Richter, Vietze, and Seifert]{PhysRevLett.103.187201}
Xiao,~R.; Fritsch,~D.; Kuz'min,~M.~D.; Koepernik,~K.; Eschrig,~H.; Richter,~M.; Vietze,~K.; Seifert,~G. Co Dimers on Hexagonal Carbon Rings Proposed as Subnanometer Magnetic Storage Bits. \emph{Phys. Rev. Lett.} \textbf{2009}, \emph{103}, 187201\relax
\mciteBstWouldAddEndPuncttrue
\mciteSetBstMidEndSepPunct{\mcitedefaultmidpunct}
{\mcitedefaultendpunct}{\mcitedefaultseppunct}\relax
\EndOfBibitem
\bibitem[Rau \latin{et~al.}(2014)Rau, Baumann, Rusponi, Donati, Stepanow, Gragnaniello, Dreiser, Piamonteze, Nolting, Gangopadhyay, Albertini, Macfarlane, Lutz, Jones, Gambardella, Heinrich, and Brune]{Rau2014}
Rau,~I.~G. \latin{et~al.}  Reaching the magnetic anisotropy limit of a 3d metal atom. \emph{Science} \textbf{2014}, \emph{344}, 988--992\relax
\mciteBstWouldAddEndPuncttrue
\mciteSetBstMidEndSepPunct{\mcitedefaultmidpunct}
{\mcitedefaultendpunct}{\mcitedefaultseppunct}\relax
\EndOfBibitem
\bibitem[Khajetoorians and Wiebe(2014)Khajetoorians, and Wiebe]{khajetoorians2014hitting}
Khajetoorians,~A.~A.; Wiebe,~J. Hitting the limit of magnetic anisotropy. \emph{Science} \textbf{2014}, \emph{344}, 976--977\relax
\mciteBstWouldAddEndPuncttrue
\mciteSetBstMidEndSepPunct{\mcitedefaultmidpunct}
{\mcitedefaultendpunct}{\mcitedefaultseppunct}\relax
\EndOfBibitem
\bibitem[Baumann \latin{et~al.}(2015)Baumann, Donati, Stepanow, Rusponi, Paul, Gangopadhyay, Rau, Pacchioni, Gragnaniello, Pivetta, Dreiser, Piamonteze, Lutz, Macfarlane, Jones, Gambardella, Heinrich, and Brune]{Baumann2015}
Baumann,~S. \latin{et~al.}  Origin of Perpendicular Magnetic Anisotropy and Large Orbital Moment in Fe Atoms on MgO. \emph{Phys. Rev. Lett.} \textbf{2015}, \emph{115}, 237202\relax
\mciteBstWouldAddEndPuncttrue
\mciteSetBstMidEndSepPunct{\mcitedefaultmidpunct}
{\mcitedefaultendpunct}{\mcitedefaultseppunct}\relax
\EndOfBibitem
\bibitem[Bouhassoune \latin{et~al.}(2016)Bouhassoune, Dias, Zimmermann, Dederichs, and Lounis]{Bouhassoune2016}
Bouhassoune,~M.; Dias,~M. d.~S.; Zimmermann,~B.; Dederichs,~P.~H.; Lounis,~S. RKKY-like contributions to the magnetic anisotropy energy: $3d$ adatoms on Pt(111) surface. \emph{Phys. Rev. B} \textbf{2016}, \emph{94}, 125402\relax
\mciteBstWouldAddEndPuncttrue
\mciteSetBstMidEndSepPunct{\mcitedefaultmidpunct}
{\mcitedefaultendpunct}{\mcitedefaultseppunct}\relax
\EndOfBibitem
\bibitem[Ibanez-Azpiroz \latin{et~al.}(2016)Ibanez-Azpiroz, dos Santos~Dias, Blugel, and Lounis]{ibanez2016zero}
Ibanez-Azpiroz,~J.; dos Santos~Dias,~M.; Blugel,~S.; Lounis,~S. Zero-point spin-fluctuations of single adatoms. \emph{Nano letters} \textbf{2016}, \emph{16}, 4305--4311\relax
\mciteBstWouldAddEndPuncttrue
\mciteSetBstMidEndSepPunct{\mcitedefaultmidpunct}
{\mcitedefaultendpunct}{\mcitedefaultseppunct}\relax
\EndOfBibitem
\bibitem[Stefan and Lounis(2018)Stefan, and Lounis]{ibanez2018spin}
Stefan; Lounis,~S. Spin-fluctuation and spin-relaxation effects of single adatoms from first principles. \emph{Journal of Physics: Condensed Matter} \textbf{2018}, \emph{30}, 343002\relax
\mciteBstWouldAddEndPuncttrue
\mciteSetBstMidEndSepPunct{\mcitedefaultmidpunct}
{\mcitedefaultendpunct}{\mcitedefaultseppunct}\relax
\EndOfBibitem
\bibitem[Bouaziz \latin{et~al.}(2020)Bouaziz, Ibaez-Azpiroz, Guimares, and Lounis]{bouaziz2020zero}
Bouaziz,~J.; Ibaez-Azpiroz,~J.; Guimares,~F.~S.; Lounis,~S. Zero-point magnetic exchange interactions. \emph{Physical review research} \textbf{2020}, \emph{2}, 043357\relax
\mciteBstWouldAddEndPuncttrue
\mciteSetBstMidEndSepPunct{\mcitedefaultmidpunct}
{\mcitedefaultendpunct}{\mcitedefaultseppunct}\relax
\EndOfBibitem
\bibitem[Donati \latin{et~al.}(2016)Donati, Rusponi, Stepanow, Wckerlin, Singha, Persichetti, Baltic, Diller, Patthey, Fernandes, Dreiser, Kummer, Nistor, Gambardella, and Brune]{Donati2016}
Donati,~F.; Rusponi,~S.; Stepanow,~S.; Wckerlin,~C.; Singha,~A.; Persichetti,~L.; Baltic,~R.; Diller,~K.; Patthey,~F.; Fernandes,~E.; Dreiser,~J.; Kummer,~K.; Nistor,~C.; Gambardella,~P.; Brune,~H. Magnetic remanence in single atoms. \emph{Science} \textbf{2016}, \emph{352}, 318--321\relax
\mciteBstWouldAddEndPuncttrue
\mciteSetBstMidEndSepPunct{\mcitedefaultmidpunct}
{\mcitedefaultendpunct}{\mcitedefaultseppunct}\relax
\EndOfBibitem
\bibitem[Donati \latin{et~al.}(2021)Donati, Pivetta, Wolf, Singha, Wackerlin, Baltic, Fernandes, De~Groot, Ahmed, Persichetti, \latin{et~al.} others]{donati2021correlation}
Donati,~F.; Pivetta,~M.; Wolf,~C.; Singha,~A.; Wackerlin,~C.; Baltic,~R.; Fernandes,~E.; De~Groot,~J.-G.; Ahmed,~S.~L.; Persichetti,~L.; others Correlation between electronic configuration and magnetic stability in dysprosium single atom magnets. \emph{Nano Letters} \textbf{2021}, \emph{21}, 8266--8273\relax
\mciteBstWouldAddEndPuncttrue
\mciteSetBstMidEndSepPunct{\mcitedefaultmidpunct}
{\mcitedefaultendpunct}{\mcitedefaultseppunct}\relax
\EndOfBibitem
\bibitem[Baumann \latin{et~al.}(2015)Baumann, Paul, Choi, Lutz, Ardavan, and Heinrich]{Baumann2015a}
Baumann,~S.; Paul,~W.; Choi,~T.; Lutz,~C.~P.; Ardavan,~A.; Heinrich,~A.~J. Electron paramagnetic resonance of individual atoms on a surface. \emph{Science} \textbf{2015}, \emph{350}, 417--420\relax
\mciteBstWouldAddEndPuncttrue
\mciteSetBstMidEndSepPunct{\mcitedefaultmidpunct}
{\mcitedefaultendpunct}{\mcitedefaultseppunct}\relax
\EndOfBibitem
\bibitem[Paul \latin{et~al.}(2017)Paul, Yang, Baumann, Romming, Choi, Lutz, and Heinrich]{Paul2017}
Paul,~W.; Yang,~K.; Baumann,~S.; Romming,~N.; Choi,~T.; Lutz,~C.; Heinrich,~A. Control of the millisecond spin lifetime of an electrically probed atom. \emph{Nat. Phys.} \textbf{2017}, \emph{13}, 403--407\relax
\mciteBstWouldAddEndPuncttrue
\mciteSetBstMidEndSepPunct{\mcitedefaultmidpunct}
{\mcitedefaultendpunct}{\mcitedefaultseppunct}\relax
\EndOfBibitem
\bibitem[Yang \latin{et~al.}(2017)Yang, Bae, Paul, Natterer, Willke, Lado, Ferr\'on, Choi, Fern\'andez-Rossier, Heinrich, and Lutz]{Yang2017}
Yang,~K.; Bae,~Y.; Paul,~W.; Natterer,~F.~D.; Willke,~P.; Lado,~J.~L.; Ferr\'on,~A.; Choi,~T.; Fern\'andez-Rossier,~J.; Heinrich,~A.~J.; Lutz,~C.~P. Engineering the Eigenstates of Coupled Spin-$1/2$ Atoms on a Surface. \emph{Phys. Rev. Lett.} \textbf{2017}, \emph{119}, 227206\relax
\mciteBstWouldAddEndPuncttrue
\mciteSetBstMidEndSepPunct{\mcitedefaultmidpunct}
{\mcitedefaultendpunct}{\mcitedefaultseppunct}\relax
\EndOfBibitem
\bibitem[Natterer \latin{et~al.}(2017)Natterer, Yang, Paul, Willke, Choi, Greber, Heinrich, and Lutz]{natterer2017reading}
Natterer,~F.~D.; Yang,~K.; Paul,~W.; Willke,~P.; Choi,~T.; Greber,~T.; Heinrich,~A.~J.; Lutz,~C.~P. Reading and writing single-atom magnets. \emph{Nature} \textbf{2017}, \emph{543}, 226--228\relax
\mciteBstWouldAddEndPuncttrue
\mciteSetBstMidEndSepPunct{\mcitedefaultmidpunct}
{\mcitedefaultendpunct}{\mcitedefaultseppunct}\relax
\EndOfBibitem
\bibitem[Forrester \latin{et~al.}(2019)Forrester, Patthey, Fernandes, Sblendorio, Brune, and Natterer]{PhysRevB.100.180405}
Forrester,~P.~R.; Patthey,~F.; Fernandes,~E.; Sblendorio,~D.~P.; Brune,~H.; Natterer,~F.~D. Quantum state manipulation of single atom magnets using the hyperfine interaction. \emph{Phys. Rev. B} \textbf{2019}, \emph{100}, 180405\relax
\mciteBstWouldAddEndPuncttrue
\mciteSetBstMidEndSepPunct{\mcitedefaultmidpunct}
{\mcitedefaultendpunct}{\mcitedefaultseppunct}\relax
\EndOfBibitem
\bibitem[Choi \latin{et~al.}(2017)Choi, Paul, Rolf-Pissarczyk, Macdonald, Natterer, Yang, Willke, Lutz, and Heinrich]{choi2017atomic}
Choi,~T.; Paul,~W.; Rolf-Pissarczyk,~S.; Macdonald,~A.~J.; Natterer,~F.~D.; Yang,~K.; Willke,~P.; Lutz,~C.~P.; Heinrich,~A.~J. Atomic-scale sensing of the magnetic dipolar field from single atoms. \emph{Nature nanotechnology} \textbf{2017}, \emph{12}, 420--424\relax
\mciteBstWouldAddEndPuncttrue
\mciteSetBstMidEndSepPunct{\mcitedefaultmidpunct}
{\mcitedefaultendpunct}{\mcitedefaultseppunct}\relax
\EndOfBibitem
\bibitem[Yang \latin{et~al.}(2021)Yang, Phark, Bae, Esat, Willke, Ardavan, Heinrich, and Lutz]{yang2021probing}
Yang,~K.; Phark,~S.-H.; Bae,~Y.; Esat,~T.; Willke,~P.; Ardavan,~A.; Heinrich,~A.~J.; Lutz,~C.~P. Probing resonating valence bond states in artificial quantum magnets. \emph{Nature communications} \textbf{2021}, \emph{12}, 1--7\relax
\mciteBstWouldAddEndPuncttrue
\mciteSetBstMidEndSepPunct{\mcitedefaultmidpunct}
{\mcitedefaultendpunct}{\mcitedefaultseppunct}\relax
\EndOfBibitem
\bibitem[Willke \latin{et~al.}(2019)Willke, Yang, Bae, Heinrich, and Lutz]{willke2019magnetic}
Willke,~P.; Yang,~K.; Bae,~Y.; Heinrich,~A.~J.; Lutz,~C.~P. Magnetic resonance imaging of single atoms on a surface. \emph{Nat. Phys.} \textbf{2019}, \emph{15}, 1005--1010\relax
\mciteBstWouldAddEndPuncttrue
\mciteSetBstMidEndSepPunct{\mcitedefaultmidpunct}
{\mcitedefaultendpunct}{\mcitedefaultseppunct}\relax
\EndOfBibitem
\bibitem[Willke \latin{et~al.}(2018)Willke, Bae, Yang, Lado, Ferr{\'o}n, Choi, Ardavan, Fern{\'a}ndez-Rossier, Heinrich, and Lutz]{Willke2018a}
Willke,~P.; Bae,~Y.; Yang,~K.; Lado,~J.~L.; Ferr{\'o}n,~A.; Choi,~T.; Ardavan,~A.; Fern{\'a}ndez-Rossier,~J.; Heinrich,~A.~J.; Lutz,~C.~P. Hyperfine interaction of individual atoms on a surface. \emph{Science} \textbf{2018}, \emph{362}, 336--339\relax
\mciteBstWouldAddEndPuncttrue
\mciteSetBstMidEndSepPunct{\mcitedefaultmidpunct}
{\mcitedefaultendpunct}{\mcitedefaultseppunct}\relax
\EndOfBibitem
\bibitem[Yang \latin{et~al.}(2018)Yang, Willke, Bae, Ferr{\'o}n, Lado, Ardavan, Fern{\'a}ndez-Rossier, Heinrich, and Lutz]{yang2018electrically}
Yang,~K.; Willke,~P.; Bae,~Y.; Ferr{\'o}n,~A.; Lado,~J.~L.; Ardavan,~A.; Fern{\'a}ndez-Rossier,~J.; Heinrich,~A.~J.; Lutz,~C.~P. Electrically controlled nuclear polarization of individual atoms. \emph{Nat. Nanotechnol.} \textbf{2018}, \emph{13}, 1120--1125\relax
\mciteBstWouldAddEndPuncttrue
\mciteSetBstMidEndSepPunct{\mcitedefaultmidpunct}
{\mcitedefaultendpunct}{\mcitedefaultseppunct}\relax
\EndOfBibitem
\bibitem[Shehada \latin{et~al.}(2021)Shehada, dos Santos~Dias, Guimar{\~a}es, Abusaa, and Lounis]{shehada2021trends}
Shehada,~S.; dos Santos~Dias,~M.; Guimar{\~a}es,~F. S.~M.; Abusaa,~M.; Lounis,~S. Trends in the hyperfine interactions of magnetic adatoms on thin insulating layers. \emph{npj Computational Materials} \textbf{2021}, \emph{7}, 1--10\relax
\mciteBstWouldAddEndPuncttrue
\mciteSetBstMidEndSepPunct{\mcitedefaultmidpunct}
{\mcitedefaultendpunct}{\mcitedefaultseppunct}\relax
\EndOfBibitem
\bibitem[Shehada \latin{et~al.}(2022)Shehada, dos Santos~Dias, Abusaa, and Lounis]{shehada2022interplay}
Shehada,~S.; dos Santos~Dias,~M.; Abusaa,~M.; Lounis,~S. Interplay of magnetic states and hyperfine fields of iron dimers on MgO (001). \emph{Journal of Physics: Condensed Matter} \textbf{2022}, \emph{34}, 385802\relax
\mciteBstWouldAddEndPuncttrue
\mciteSetBstMidEndSepPunct{\mcitedefaultmidpunct}
{\mcitedefaultendpunct}{\mcitedefaultseppunct}\relax
\EndOfBibitem
\bibitem[Kim \latin{et~al.}(2022)Kim, Noh, Chen, Donati, Heinrich, Wolf, and Bae]{kim2022anisotropic}
Kim,~J.; Noh,~K.; Chen,~Y.; Donati,~F.; Heinrich,~A.~J.; Wolf,~C.; Bae,~Y. Anisotropic hyperfine interaction of surface-adsorbed single atoms. \emph{Nano Letters} \textbf{2022}, \emph{22}, 9766--9772\relax
\mciteBstWouldAddEndPuncttrue
\mciteSetBstMidEndSepPunct{\mcitedefaultmidpunct}
{\mcitedefaultendpunct}{\mcitedefaultseppunct}\relax
\EndOfBibitem
\bibitem[Farinacci \latin{et~al.}(2022)Farinacci, Veldman, Willke, and Otte]{farinacci2022experimental}
Farinacci,~L.; Veldman,~L.~M.; Willke,~P.; Otte,~S. Experimental Determination of a Single Atom Ground State Orbital through Hyperfine Anisotropy. \emph{Nano Letters} \textbf{2022}, \emph{22}, 8470--8474\relax
\mciteBstWouldAddEndPuncttrue
\mciteSetBstMidEndSepPunct{\mcitedefaultmidpunct}
{\mcitedefaultendpunct}{\mcitedefaultseppunct}\relax
\EndOfBibitem
\bibitem[Zhang \latin{et~al.}(2022)Zhang, Wolf, Wang, Aubin, Bilgeri, Willke, Heinrich, and Choi]{zhang2022electron}
Zhang,~X.; Wolf,~C.; Wang,~Y.; Aubin,~H.; Bilgeri,~T.; Willke,~P.; Heinrich,~A.~J.; Choi,~T. Electron spin resonance of single iron phthalocyanine molecules and role of their non-localized spins in magnetic interactions. \emph{Nature Chemistry} \textbf{2022}, \emph{14}, 59--65\relax
\mciteBstWouldAddEndPuncttrue
\mciteSetBstMidEndSepPunct{\mcitedefaultmidpunct}
{\mcitedefaultendpunct}{\mcitedefaultseppunct}\relax
\EndOfBibitem
\bibitem[Willke \latin{et~al.}(2021)Willke, Bilgeri, Zhang, Wang, Wolf, Aubin, Heinrich, and Choi]{willke2021coherent}
Willke,~P.; Bilgeri,~T.; Zhang,~X.; Wang,~Y.; Wolf,~C.; Aubin,~H.; Heinrich,~A.; Choi,~T. Coherent Spin Control of Single Molecules on a Surface. \emph{ACS nano} \textbf{2021}, \emph{15}, 17959--17965\relax
\mciteBstWouldAddEndPuncttrue
\mciteSetBstMidEndSepPunct{\mcitedefaultmidpunct}
{\mcitedefaultendpunct}{\mcitedefaultseppunct}\relax
\EndOfBibitem
\bibitem[Kim \latin{et~al.}(2021)Kim, Jang, Bui, Choi, Wolf, Delgado, Chen, Krylov, Lee, Yoon, \latin{et~al.} others]{kim2021spin}
Kim,~J.; Jang,~W.-j.; Bui,~T.~H.; Choi,~D.-J.; Wolf,~C.; Delgado,~F.; Chen,~Y.; Krylov,~D.; Lee,~S.; Yoon,~S.; others Spin resonance amplitude and frequency of a single atom on a surface in a vector magnetic field. \emph{Physical Review B} \textbf{2021}, \emph{104}, 174408\relax
\mciteBstWouldAddEndPuncttrue
\mciteSetBstMidEndSepPunct{\mcitedefaultmidpunct}
{\mcitedefaultendpunct}{\mcitedefaultseppunct}\relax
\EndOfBibitem
\bibitem[Steinbrecher \latin{et~al.}(2021)Steinbrecher, van Weerdenburg, Walraven, van Mullekom, Gerritsen, Natterer, Badrtdinov, Rudenko, Mazurenko, Katsnelson, van~der Avoird, Groenenboom, and Khajetoorians]{PhysRevB.103.155405}
Steinbrecher,~M.; van Weerdenburg,~W. M.~J.; Walraven,~E.~F.; van Mullekom,~N. P.~E.; Gerritsen,~J.~W.; Natterer,~F.~D.; Badrtdinov,~D.~I.; Rudenko,~A.~N.; Mazurenko,~V.~V.; Katsnelson,~M.~I.; van~der Avoird,~A.; Groenenboom,~G.~C.; Khajetoorians,~A.~A. Quantifying the interplay between fine structure and geometry of an individual molecule on a surface. \emph{Phys. Rev. B} \textbf{2021}, \emph{103}, 155405\relax
\mciteBstWouldAddEndPuncttrue
\mciteSetBstMidEndSepPunct{\mcitedefaultmidpunct}
{\mcitedefaultendpunct}{\mcitedefaultseppunct}\relax
\EndOfBibitem
\bibitem[Singha \latin{et~al.}(2021)Singha, Sostina, Wolf, Ahmed, Krylov, Colazzo, Gargiani, Agrestini, Noh, Park, \latin{et~al.} others]{singha2021mapping}
Singha,~A.; Sostina,~D.; Wolf,~C.; Ahmed,~S.~L.; Krylov,~D.; Colazzo,~L.; Gargiani,~P.; Agrestini,~S.; Noh,~W.-S.; Park,~J.-H.; others Mapping Orbital-Resolved Magnetism in Single Lanthanide Atoms. \emph{ACS nano} \textbf{2021}, \emph{15}, 16162--16171\relax
\mciteBstWouldAddEndPuncttrue
\mciteSetBstMidEndSepPunct{\mcitedefaultmidpunct}
{\mcitedefaultendpunct}{\mcitedefaultseppunct}\relax
\EndOfBibitem
\bibitem[Singha \latin{et~al.}(2021)Singha, Willke, Bilgeri, Zhang, Brune, Donati, Heinrich, and Choi]{singha2021engineering}
Singha,~A.; Willke,~P.; Bilgeri,~T.; Zhang,~X.; Brune,~H.; Donati,~F.; Heinrich,~A.~J.; Choi,~T. Engineering atomic-scale magnetic fields by dysprosium single atom magnets. \emph{Nature Communications} \textbf{2021}, \emph{12}, 1--6\relax
\mciteBstWouldAddEndPuncttrue
\mciteSetBstMidEndSepPunct{\mcitedefaultmidpunct}
{\mcitedefaultendpunct}{\mcitedefaultseppunct}\relax
\EndOfBibitem
\bibitem[Yang \latin{et~al.}(2019)Yang, Paul, Phark, Willke, Bae, Choi, Esat, Ardavan, Heinrich, and Lutz]{Yang2019a}
Yang,~K.; Paul,~W.; Phark,~S.-H.; Willke,~P.; Bae,~Y.; Choi,~T.; Esat,~T.; Ardavan,~A.; Heinrich,~A.~J.; Lutz,~C.~P. Coherent spin manipulation of individual atoms on a surface. \emph{Science} \textbf{2019}, \emph{366}, 509--512\relax
\mciteBstWouldAddEndPuncttrue
\mciteSetBstMidEndSepPunct{\mcitedefaultmidpunct}
{\mcitedefaultendpunct}{\mcitedefaultseppunct}\relax
\EndOfBibitem
\bibitem[Kovarik \latin{et~al.}(2022)Kovarik, Robles, Schlitz, Seifert, Lorente, Gambardella, and Stepanow]{kovarik2022electron}
Kovarik,~S.; Robles,~R.; Schlitz,~R.; Seifert,~T.~S.; Lorente,~N.; Gambardella,~P.; Stepanow,~S. Electron paramagnetic resonance of alkali metal atoms and dimers on ultrathin MgO. \emph{Nano Letters} \textbf{2022}, \relax
\mciteBstWouldAddEndPunctfalse
\mciteSetBstMidEndSepPunct{\mcitedefaultmidpunct}
{}{\mcitedefaultseppunct}\relax
\EndOfBibitem
\bibitem[Garai-Marin \latin{et~al.}(2023)Garai-Marin, dos Santos~Dias, Lounis, Ibaez-Azpiroz, and Eiguren]{garai2023microscopic}
Garai-Marin,~H.; dos Santos~Dias,~M.; Lounis,~S.; Ibaez-Azpiroz,~J.; Eiguren,~A. Microscopic theory of spin relaxation of a single Fe adatom coupled to substrate vibrations. \emph{Physical Review B} \textbf{2023}, \emph{107}, 144417\relax
\mciteBstWouldAddEndPuncttrue
\mciteSetBstMidEndSepPunct{\mcitedefaultmidpunct}
{\mcitedefaultendpunct}{\mcitedefaultseppunct}\relax
\EndOfBibitem
\bibitem[Ou \latin{et~al.}(2015)Ou, Wang, Fan, Li, and Wu]{ou2015giant}
Ou,~X.; Wang,~H.; Fan,~F.; Li,~Z.; Wu,~H. Giant magnetic anisotropy of Co, Ru, and Os adatoms on MgO (001) surface. \emph{Physical Review Letters} \textbf{2015}, \emph{115}, 257201\relax
\mciteBstWouldAddEndPuncttrue
\mciteSetBstMidEndSepPunct{\mcitedefaultmidpunct}
{\mcitedefaultendpunct}{\mcitedefaultseppunct}\relax
\EndOfBibitem
\bibitem[Wang \latin{et~al.}(1993)Wang, Wu, and Freeman]{wang1993first}
Wang,~D.-s.; Wu,~R.; Freeman,~A. First-principles theory of surface magnetocrystalline anisotropy and the diatomic-pair model. \emph{Physical Review B} \textbf{1993}, \emph{47}, 14932\relax
\mciteBstWouldAddEndPuncttrue
\mciteSetBstMidEndSepPunct{\mcitedefaultmidpunct}
{\mcitedefaultendpunct}{\mcitedefaultseppunct}\relax
\EndOfBibitem
\bibitem[Brooks(1940)]{brooks1940ferromagnetic}
Brooks,~H. Ferromagnetic anisotropy and the itinerant electron model. \emph{Physical Review} \textbf{1940}, \emph{58}, 909\relax
\mciteBstWouldAddEndPuncttrue
\mciteSetBstMidEndSepPunct{\mcitedefaultmidpunct}
{\mcitedefaultendpunct}{\mcitedefaultseppunct}\relax
\EndOfBibitem
\bibitem[Bruno(1989)]{bruno1989tight}
Bruno,~P. Tight-binding approach to the orbital magnetic moment and magnetocrystalline anisotropy of transition-metal monolayers. \emph{Physical Review B} \textbf{1989}, \emph{39}, 865\relax
\mciteBstWouldAddEndPuncttrue
\mciteSetBstMidEndSepPunct{\mcitedefaultmidpunct}
{\mcitedefaultendpunct}{\mcitedefaultseppunct}\relax
\EndOfBibitem
\bibitem[Dai \latin{et~al.}(2008)Dai, Xiang, and Whangbo]{dai2008effects}
Dai,~D.; Xiang,~H.; Whangbo,~M.-H. Effects of spin-orbit coupling on magnetic properties of discrete and extended magnetic systems. \emph{Journal of computational chemistry} \textbf{2008}, \emph{29}, 2187--2209\relax
\mciteBstWouldAddEndPuncttrue
\mciteSetBstMidEndSepPunct{\mcitedefaultmidpunct}
{\mcitedefaultendpunct}{\mcitedefaultseppunct}\relax
\EndOfBibitem
\bibitem[Sakurai and Commins(1995)Sakurai, and Commins]{sakurai1995modern}
Sakurai,~J.~J.; Commins,~E.~D. Modern quantum mechanics, revised edition. 1995\relax
\mciteBstWouldAddEndPuncttrue
\mciteSetBstMidEndSepPunct{\mcitedefaultmidpunct}
{\mcitedefaultendpunct}{\mcitedefaultseppunct}\relax
\EndOfBibitem
\bibitem[Lu \latin{et~al.}(2013)Lu, Zuo, Feng, and Zhou]{lu2013magnetic}
Lu,~Y.; Zuo,~X.; Feng,~M.; Zhou,~T. Magnetic anisotropy in the boron nitride monolayer doped by 3d transitional metal substitutes at boron-site. \emph{Journal of Applied Physics} \textbf{2013}, \emph{113}, 17C304\relax
\mciteBstWouldAddEndPuncttrue
\mciteSetBstMidEndSepPunct{\mcitedefaultmidpunct}
{\mcitedefaultendpunct}{\mcitedefaultseppunct}\relax
\EndOfBibitem
\bibitem[Shao \latin{et~al.}(2014)Shao, Feng, and Zuo]{shao2014carrier}
Shao,~B.; Feng,~M.; Zuo,~X. Carrier-dependent magnetic anisotropy of cobalt doped titanium dioxide. \emph{Scientific Reports} \textbf{2014}, \emph{4}, 1--6\relax
\mciteBstWouldAddEndPuncttrue
\mciteSetBstMidEndSepPunct{\mcitedefaultmidpunct}
{\mcitedefaultendpunct}{\mcitedefaultseppunct}\relax
\EndOfBibitem
\bibitem[Van Der~Laan(1991)]{van1991m2}
Van Der~Laan,~G. M2, 3 absorption spectroscopy of 3d transition-metal compounds. \emph{Journal of Physics: Condensed Matter} \textbf{1991}, \emph{3}, 7443\relax
\mciteBstWouldAddEndPuncttrue
\mciteSetBstMidEndSepPunct{\mcitedefaultmidpunct}
{\mcitedefaultendpunct}{\mcitedefaultseppunct}\relax
\EndOfBibitem
\bibitem[Vanderbilt(1990)]{vanderbilt1990soft}
Vanderbilt,~D. Soft self-consistent pseudopotentials in a generalized eigenvalue formalism. \emph{Physical review B} \textbf{1990}, \emph{41}, 7892\relax
\mciteBstWouldAddEndPuncttrue
\mciteSetBstMidEndSepPunct{\mcitedefaultmidpunct}
{\mcitedefaultendpunct}{\mcitedefaultseppunct}\relax
\EndOfBibitem
\bibitem[Dal~Corso and Conte(2005)Dal~Corso, and Conte]{dal2005spin}
Dal~Corso,~A.; Conte,~A.~M. Spin-orbit coupling with ultrasoft pseudopotentials: Application to Au and Pt. \emph{Physical Review B} \textbf{2005}, \emph{71}, 115106\relax
\mciteBstWouldAddEndPuncttrue
\mciteSetBstMidEndSepPunct{\mcitedefaultmidpunct}
{\mcitedefaultendpunct}{\mcitedefaultseppunct}\relax
\EndOfBibitem
\bibitem[Dal~Corso(2014)]{dal2014pseudopotentials}
Dal~Corso,~A. Pseudopotentials periodic table: From H to Pu. \emph{Computational Materials Science} \textbf{2014}, \emph{95}, 337--350\relax
\mciteBstWouldAddEndPuncttrue
\mciteSetBstMidEndSepPunct{\mcitedefaultmidpunct}
{\mcitedefaultendpunct}{\mcitedefaultseppunct}\relax
\EndOfBibitem
\bibitem[Perdew and Zunger(1981)Perdew, and Zunger]{PhysRevB.23.5048}
Perdew,~J.~P.; Zunger,~A. Self-interaction correction to density-functional approximations for many-electron systems. \emph{Phys. Rev. B} \textbf{1981}, \emph{23}, 5048--5079\relax
\mciteBstWouldAddEndPuncttrue
\mciteSetBstMidEndSepPunct{\mcitedefaultmidpunct}
{\mcitedefaultendpunct}{\mcitedefaultseppunct}\relax
\EndOfBibitem
\bibitem[Liechtenstein \latin{et~al.}(1995)Liechtenstein, Anisimov, and Zaanen]{liechtenstein1995density}
Liechtenstein,~A.; Anisimov,~V.~I.; Zaanen,~J. Density-functional theory and strong interactions: Orbital ordering in Mott-Hubbard insulators. \emph{Physical Review B} \textbf{1995}, \emph{52}, R5467\relax
\mciteBstWouldAddEndPuncttrue
\mciteSetBstMidEndSepPunct{\mcitedefaultmidpunct}
{\mcitedefaultendpunct}{\mcitedefaultseppunct}\relax
\EndOfBibitem
\bibitem[Ma and Dudarev(2015)Ma, and Dudarev]{PhysRevB.91.054420}
Ma,~P.-W.; Dudarev,~S.~L. Constrained density functional for noncollinear magnetism. \emph{Phys. Rev. B} \textbf{2015}, \emph{91}, 054420\relax
\mciteBstWouldAddEndPuncttrue
\mciteSetBstMidEndSepPunct{\mcitedefaultmidpunct}
{\mcitedefaultendpunct}{\mcitedefaultseppunct}\relax
\EndOfBibitem
\bibitem[{J\"{u}lich Supercomputing Centre}(2018)]{jureca}
{J\"{u}lich Supercomputing Centre} {JURECA: Modular supercomputer at J\"{u}lich Supercomputing Centre}. \emph{JLSRF} \textbf{2018}, \emph{4}\relax
\mciteBstWouldAddEndPuncttrue
\mciteSetBstMidEndSepPunct{\mcitedefaultmidpunct}
{\mcitedefaultendpunct}{\mcitedefaultseppunct}\relax
\EndOfBibitem
\end{mcitethebibliography}

\end{document}